\documentclass{PoS}

\newcommand\psib{{\overline{\psi}}}
\newcommand\zb{{\overline{z}}}
\newcommand\ub{{\overline{u}}}
\newcommand\db{{\overline{d}}}

\usepackage{amsmath}

\title{A new computational approach to lattice quantum field theories}

\ShortTitle{A new computational approach to LQFT}

\author{\speaker{Shailesh Chandrasekharan}\thanks{This work was supported in part by a DOE grant DE-FG02-05ER41368.}\\
        Department of Physics, Duke University, Durham, NC 27708-0305, USA\\
        E-mail: \email{sch@phy.duke.edu}}

\abstract{Developments in algorithms over the past decade suggest that there is a new computational approach to a class of quantum field theories. This approach is based on rewriting the partition function in a representation similar to the world-line representation and hence we shall call it the ``WL-approach''. This approach is likely to be more powerful than the conventional approach in some regions of parameter space, especially in the presence of chemical potentials or massless fermions. While world-line representations are natural in the Hamiltonian formulation, they can also be constructed directly in Euclidean space. We first describe the approach and its advantages by considering the classical XY model in the presence of a chemical potential. We then argue that, $CP^{N-1}$ models, models of pions on the lattice and the lattice massless Thirring model, can all be formulated and solved using the WL-approach. In particular, we discover that the WL-approach to the Thirring model leads to a novel determinantal Monte-Carlo algorithm which we call the ``dynamical-bag'' algorithm. Finally, we argue that a simple extension of the WL-approach to gauge theories leads to a world-sheet, ``WS-approach'', in Abelian Lattice Gauge theory.}

\FullConference{The XXVI International Symposium on Lattice Field Theory\\
		 July 14-19 2008\\
		 Williamsburg, Virginia, USA}

\begin{document}

\section{Introduction}

Monte Carlo methods have tremendously improved our understanding of a variety of strongly interacting quantum field theories. Beginning with the discovery of ``cluster algorithms'' for classical spin systems \cite{PhysRevLett.58.86,PhysRevLett.62.361}, and the ``loop algorithms'' for quantum spin systems  \cite{PhysRevLett.70.875} the search for efficient algorithms has been an important area of research over the past two decades. Although many algorithmic improvements have been achieved for QCD-like problems \cite{Orginos:2006en}, it has been especially difficult to find good algorithms for lattice field theory problems in the presence of a chemical potential or when the microscopic theory contains strongly interacting massless fermions. However, recently the landscape has begun to change somewhat. The progress over the past decade, especially in studies of quantum spin systems and strong coupling lattice gauge theories, suggests the existence of a new approach to a class of lattice field theories. In this talk I will outline the ideas behind the new approach by discussing a few examples.

One essential feature of the new approach is to formulate lattice field theories in representations that are similar to the world-line representation. Such representations arise naturally in the Hamiltonian formulation but are not usually used in the Lagrangian formulation. As we will see below, they also arise naturally in Euclidean lattice field theory when one thinks about the strong coupling (or high temperature) expansions. In certain cases the representations are indeed the well-known world-line representations, in others they look different but share many properties of the world line representations. For this reason we will refer to the new approach generically as the ``WL-approach''. As we will see below one of the advantages of the world-line representations is that there are no new sign problems because of adding a chemical potential unlike the conventional approach. On the other hand, the world-line representations are not friendly for local Monte Carlo methods due to the presence of constraints. The loop cluster algorithm was the first successful algorithm that was able solve the constrained problem efficiently. However, this algorithm was rather restrictive and it was difficult to maintain the efficiency in the presence of different types of couplings especially with a chemical potential. Recently, it was realized that the loop algorithm is just one example of a more general class of algorithms referred to as ``worm algorithms'' \cite{PhysRevLett.87.160601} or ``directed-loop algorithms'' \cite{PhysRevE.66.046701} or ``directed-path algorithm'' \cite{Adams:2003cc}. We will refer to these slightly different algorithms together as ``worm-algorithms'' based on the pioneering work of the first authors Prokof\'ev and Svustinov. These algorithms introduce a defect in the constraints and sample the defect space until the defect naturally eliminates itself. The sampled defect space is itself a measure of the two point correlation function. When combined with the worm-algorithm, the WL-approach to lattice field theories have been found to be as efficient as the original cluster algorithms for classical spin models \cite{PhysRevLett.87.160601,Wolff:2008km}. On the other hand the WL-approach seem to be more widely applicable. As we will discuss below, the worm-algorithm can in principle be applied to $CP^{N-1}$ models which cannot be solved using the conventional cluster algorithms.

As with any Monte-Carlo method, the WL-approach can fail due to sign problems. Sign problems can arise due to frustrating interactions or in the presence of fermions. In two dimensions fermionic sign problems can sometimes be solved in the world line representation since fermions resemble hard-core bosons \cite{Gattringer:2007em,Wolff:2007ip}. Thus, WL-approach should be applicable to these problems. Solutions to fermionic sign problems also emerge in strongly coupled lattice gauge theories with staggered fermions. In this case configurations naturally resemble world-line representations where fermions are confined into bosons. When combined with the efficient directed-path algorithm \cite{Adams:2003cc} this provides a new and powerful WL-approach for some scalar field theories which include those that are invariant under symmetries like $SU(2)\times U(1)$ and $SU(2)\times SU(2)\times U(1)$ \cite{Chandrasekharan:2006tz,Chandrasekharan:2007up,Cecile:2007dv}. These theories, which are of interest in the context of understanding QCD-like theories, could not be solved efficiently with conventional approaches.

In higher dimensions solving fermionic sign problems is difficult, however the WL-approach suggests new solutions in many cases \cite{PhysRevLett.83.3116,Chandrasekharan:1999vz,Chandrasekharan:2002vk,yoo:201309}. If the sign problems can be solved in the WL-approach, then one can combine worm-algorithms with other conventional algorithms to solve the problem. Below we will discuss one such algorithm for the massless lattice Thirring model in any dimension. In the WL-approach the partition function is rewritten such that it resembles the well known bag-model of QCD where the bags are dynamically determined objects. Inside the bag the fermions are free, while outside they are confined. For this reason we call this new fermion algorithm the ``dynamical-bag algorithm''.

The basic ideas of the WL-approach are also applicable to gauge theories. Here one needs to rewrite the problem in a representation of world-sheets giving us a new ``WS-approach'' where a worm-type algorithm involves cutting and reconnecting world sheets. Although very little is known about these surface-cluster algorithms, the ideas seem promising. Below we will show some results which suggest that it may be possible to measure large Wilson loops with little effort in the confined phase in Abelian lattice gauge theory using a WS-approach.

\section{World-Line Approach: XY Model with a Chemical Potential}

In this section we will illustrate the WL-approach using the example of the XY model. To make things interesting we will introduce a chemical potential that couples to the $U(1)$ charge of the particles. We begin with the action of the lattice field theory, in the conventional formulation, which is given by
\begin{equation}
S = - \beta
\sum_{x,\alpha}
\Bigg\{\cos\Big([\phi_x-\phi_{x+\alpha}]-i\mu\delta_{\alpha,t}\Big) \Bigg\}
\end{equation}
where $x$ is a point on a $d+1$-dimensional hypercubic lattice, $\alpha=t,1,2..,d$ represents the direction, $\beta$ is the inverse coupling of the theory and $\exp(i\phi_x)$ is the non-linear $U(1)$ bosonic field. Clearly, when $\mu\neq 0$ the action is complex and the path integral suffers from a sign problem and one cannot use the conventional Monte Carlo methods. In particular the Wolff-cluster algorithm \cite{PhysRevLett.62.361} is no longer applicable.

\begin{figure}[t]
\begin{center}
\includegraphics[width=0.6\textwidth]{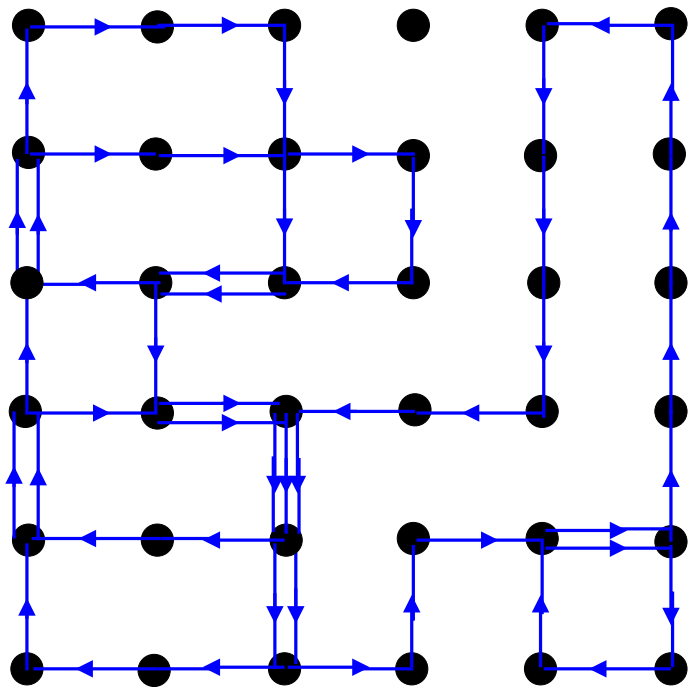}
\end{center}
\caption{\label{XYconf} An illustration of the world-line configuration for the XY-model. }
\end{figure}

In order to construct the WL-approach we rewrite the partition function using the strong coupling (or high temperature) expansion to all orders. Using the identity
\begin{equation}
\label{modbf}
\exp\Big\{\beta \cos(\phi)\Big\}
= \sum_{k=-\infty}^\infty I_k\Big(\beta\Big)\mathrm{e}^{ik\phi},
\end{equation}
on every bond and integrating over the original angle variables the partition function can be written in terms of integer bond variables $k_{x,\alpha}$ and one gets
\begin{equation}
Z = \sum_{[k_{x,\alpha}]} \ \ \prod_{[x,\alpha]} 
\Bigg\{\mathrm{e}^{\mu \delta_{\alpha,t} k_{x,\alpha}} \ 
I_{k_{x,\alpha}}\Big(\beta\Big) \  
\delta\Big(\sum_\alpha [k_{x,\alpha} - k_{x-\alpha,\alpha}]\Big)\Bigg\}.
\label{currloop}
\end{equation}
Here $I_k(\beta)$ is the modified Bessel function of the first kind. The delta function in the above expression shows that the bond variables satisfy a local constraint at each lattice site $x$ which is nothing but the current conservation relation. Thus, the partition function of the non-linear sigma model has been rewritten as a sum over configurations of current-loops. Figure \ref{fig1} illustrates one such configuration in two dimensions. Importantly, this representation does not suffer from a sign problem even at non-zero $\mu$. The sign problem has been traded for a constraint condition which is not easily satisfied by a local change. However, developments over the past decade have shown that there are efficient non-local updates for such constrained problems \cite{PhysRevLett.87.160601,PhysRevE.66.046701,Adams:2003cc,boninsegni:070601}. 

Is this approach general? In other words can the partition function of a lattice field theory be written in a world-line representation and then a worm algorithm be used to solve the problem? We will argue below that the answer is ``yes'' in many cases. This was recently shown to be true for the Ising model \cite{Wolff:2008km}. In the next section we will argue that the $CP^{N-1}$ model can also be solved in this approach.

\section{World-line Approach: $CP^{N-1}$ Model}

In this section we will construct the world-line representation of the $CP^{N-1}$ model as a non-trivial application of the WL-approach. One may recall that conventional cluster algorithms do not work for these models even in the absence of a chemical potential. $CP^{N-1}$ models were recently formulated as the low energy effective theory of an $SU(N)$-symmetric quantum-spin system in the Hamiltonian (D-theory) formulation \cite{Beard:2006mq}. The spin system was formulated in the world-line formulation, and solved using the loop-cluster algorithm. Thus, the first successful efficient cluster algorithm for $CP^{N-1}$ models emerged in the WL-approach. Here we will argue that a world-line representation can also be constructed for the conventional model directly in Euclidean space.

The $CP^{N-1}$ model is conventionally written in terms of $N$ component complex vector $z^a_i, a=1,2,..,N$ at each lattice site $i$ with the constraint $\sum_a |z^a_i|^2 = 1$ \cite{Eichenherr:1978qa,D'Adda:1978uc}. The action of the model is given by
\begin{equation}
\label{act}
S = -\beta \sum_{<ij>} (\zb^a_i z^b_i) (\zb^b_j z^a_j)
\end{equation}
where $<ij>$ denotes the bond connecting the nearest neighbor sites $i$ and $j$. The action is invariant under global $SU(N)$ transformations $z_i \rightarrow U z_i$ and $U(1)$ gauge transformations $z_i \rightarrow \exp(i\phi_i) z_i$. Since the gauge field is not dynamical in the theory, the model is equivalent to strongly coupled $N$-flavor scalar QED. Let us now derive the world-line representation.

Consider the partition function of the model,
\begin{equation}
\label{pf}
Z = \int \prod_i [dz_i] \exp(\beta \sum_{<ij>} (\zb^a_i z^b_i) (\zb^b_j z^a_j))
\end{equation}
where $[dz_i]$ is the integral over $2N$-dimensional unit vectors. The partition function can be rewritten as
\begin{equation}
Z = \int \prod_i [dz_i] \prod_{<ij>} \prod_{<ab>}
\sum_{[n^{ab}_{ij}]} \frac{\beta^{n^{ab}_{ij}}}{n^{ab}_{ij}!}
(\zb^a_i z^b_i)^{n^{ab}_{ij}} (\zb^b_j z^a_j)^{n^{ab}_{ij}}
\end{equation}
where each bond $<ij>$ contains an $N\times N$ matrix of non-negative integers $n_{ij} \equiv n^{ab}_{ij}$ which determines the power of $(\zb^a_i z^b_i)$ and $(\zb^b_j z^a_j)$. Now one can integrate over $[dz_i]$ using the identity
\begin{equation}
\int [dz] (\zb_1)^{k_1}(z_1)^{l_1}...(\zb_N)^{k_N}(z_N)^{l_N}
= \delta_{k_1,l_1}\delta_{k_2,l_2}...\delta_{k_N,l_N}\ 
\frac{2\pi^N k_1! k_2! k_3!...k_N!}{(k_1+k_2+...+k_N+N-1)!}.
\end{equation}
If we define $k \equiv(k_1,k_2,...,k_N)$ as an $N$-vector and
\begin{equation}
I(k) = \frac{2\pi^N k_1! k_2! k_3!...k_N!}{(k_1+k_2+...+k_N+N-1)!},
\end{equation}
the partition function can be written compactly as
\begin{equation}
Z = \sum_{[n_{ij}]} \Bigg(\prod_i I(q^i)\Bigg) \delta_{q^i,p^i} 
\Bigg(\prod_{<ij>} \prod_{ab} \frac{\beta^{n^{ab}_{ij}}}{n^{ab}_{ij}!}\Bigg)
\end{equation}
where
\begin{equation}
q^i_a = \sum_\mu \sum_b \Big\{n^{ab}_{i (i+\mu)} + n^{ba}_{(i-\mu) i}\Big\},\ 
p^i_a = \sum_\mu \sum_b \Big\{n^{ba}_{i (i+\mu)} + n^{ab}_{(i-\mu) i}\Big\}
\end{equation}
Thus, the partition function has been written in terms of the constrained bond matrices $n_{ij}$. These constrains encode the conserved currents of the $CP^{N-1}$ model and hence we call this the world-line representation. Using the ideas developed in \cite{Adams:2003cc} it should now be possible to update the constrained system. This has not yet been accomplished but is an interesting research project for the future.

\section{Bosons as Fermionic Composites}

Scalar field theories are naturally formulated using scalar fields. However, all scalar particles discovered in nature until now are bound fermionic composites. Interestingly, we have discovered recently that it is also computationally simpler to formulate certain scalar lattice field theories from a microscopic model of fermions such that the bosons arise as fermionic composites. Composite fermion models of scalar field theories can be constructed easily as strongly coupled lattice gauge theories. They lead to a novel world-line approach where the fermionic nature restricts the number of allowed scalar particles on a given lattice site. The underlying idea is very similar to the D-theory formulations of field theories \cite{Brower:2003vy}, however the formulations discussed below occur directly in Euclidean space and often it is easier to design worm-type algorithms in this approach. Let us illustrate this with two examples.

\begin{figure}[t]
\begin{center}
\includegraphics[width=0.6\textwidth]{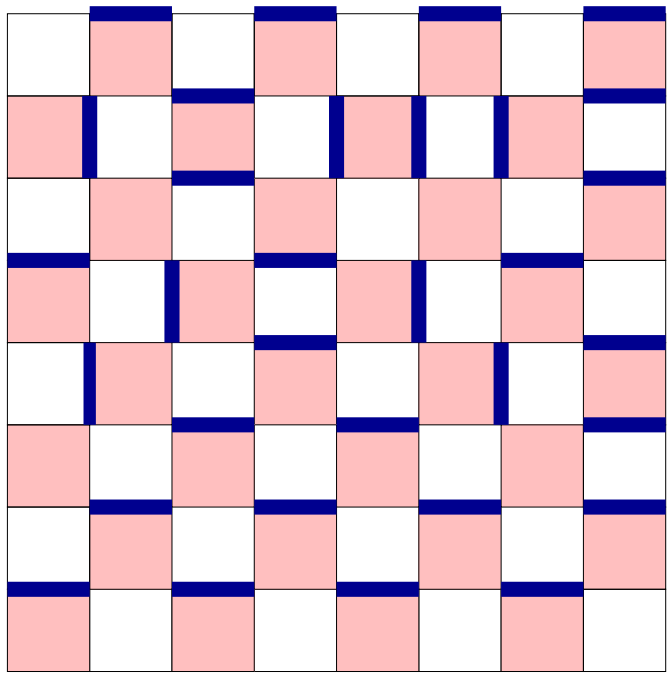}
\end{center}
\caption{\label{fXYconf} An illustration of a closed pack dimer configuration in two dimension. In $d+1$-dimensions statistical mechanics of such configurations naturally lead to the physics of the $d$-dimensional XY model.}
\end{figure}

First consider the $XY$ model. A simple $XY$ model of composite fermions on a $d+1$ dimensional hypercubic lattice is given by
\begin{equation}
S = -\sum_{x,i=1,2..,d} \psib_x\psi_x\psib_{x+i}\psi_{x+i} - T \sum_{x} \psib_x\psi_x\psib_{x+t}\psi_{x+t}
\label{fxy}
\end{equation}
where $\psi_x,\psib_x$ are two Grassmann valued lattice fields. This theory has an exact global $U(1)$ symmetry where $\psi_x \rightarrow e^{i\sigma_x\theta}\psi_x$ and $\psib_x \rightarrow e^{i\sigma_x\theta}\psib_x$ where $\sigma_x$ is $+1$ on even sites and $-1$ on odd sites. Note that one of the dimensions (referred to as $t$ here) has been singled out and acts like a fictitious temperature. If we study this theory on a $L^d\times L_t$ lattice with $L_t$ fixed, then by tuning $T$ one obtains the physics of the conventional $XY$ model in $d$ dimensions through dimensional reduction.

It is possible to integrate out the Grassmann variables and write the partition function as a sum over all possible closed packed dimer configurations. In particular every dimer in the temporal direction carries a weight $T$. Thus, the statistical mechanics of closed packed dimers on a cubic lattice in $d+1$-dimensions leads naturally to the physics of the $XY$ model. A configuration of a closed pack dimers is shown in Fig.~\ref{fXYconf}. Clearly these configurations are much simpler to represent on a computer than the world-line configurations of the conventional $XY$ model shown in Fig.~\ref{XYconf}. Another advantage of this representation of the $XY$ model is that it can be updated very efficiently with a worm-type algorithm which is much simpler to code than the Wolff-cluster algorithm \cite{PhysRevLett.62.361}. It has been shown that this model and its variants do indeed reproduce the expected $KT$ physics in $d=2$ \cite{Chandrasekharan:2003qv} and $XY$ universal critical behavior in $d=3$ \cite{Chandrasekharan:2003im}. Interestingly, in $1+1$ dimensions with $T=1$ and $L_t=L$, the model leads to a non-compact version of the $XY$ model and so contains no vortices and evades a KT transition \cite{Cecile:2008nb}. Thus, the fermionic representation also allows us to formulate naturally a non-compact $U(1)$ field theory.

Second we focus on an $SU(2)\times SU(2)\times U(1)$ model of composite fermions. This is an interesting toy model for pions of two flavor QCD. The action of the model is given by
\begin{equation}
S = -\sum_{x,i=1,2..,d} \mathrm{Tr}[\Sigma_x \ \Sigma_{x+i}] - T \sum_{x} \mathrm{Tr}[\Sigma_x \ \Sigma_{x+t}] - c \sum_x \mathrm{det} \Sigma_x
\end{equation}
where
\begin{equation}
\Sigma_x \equiv \left(\begin{array}{c} u_x \cr d_x \end{array} \right) 
\left(\begin{array}{cc} \ub_x & \ \ \db_x \end{array} \right) =
\left(\begin{array}{cc} u_x\ub_x & \ \ \ u_x\db_x \cr d_x\ub_x & \ \ \ d_x\db_x\end{array} \right),
\end{equation}
is a $2\times 2$ matrix made up fermionic bilinears. Here $u_x,\ub_x,d_x$ and $\db_x$ are four independent Grassmann variables defined on each site. When $c=0$, the action is symmetric under the $SU(2)\times  SUS(2) \times U(1)$ chiral transformations
\begin{eqnarray}
&& \Sigma_x \rightarrow L \Sigma_x R^\dagger \mathrm{e}^{i\phi} \hspace{1in}
\mbox{for $x$ even} \\ \nonumber
&& \Sigma_x \rightarrow R \Sigma_x L^\dagger \mathrm{e}^{-i\phi} \hspace{1in} 
\mbox{for $x$ odd} \\ \nonumber
\end{eqnarray}
When $c\neq 0$, the symmetry under the $U(1)$ subgroup is broken explicitly. Hence the parameter $c$ behaves like the anomaly. 

\begin{figure}[t]
\begin{center}
\includegraphics[width=0.6\textwidth]{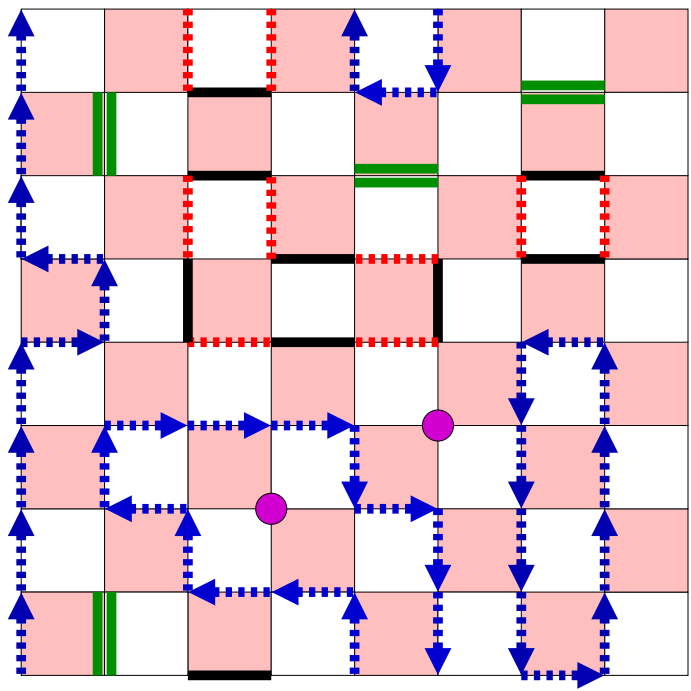}
\end{center}
\caption{\label{fPIconf} An illustration of a world-line configuration for an $SU(2)\times SU(2) \times U(1)$ scalar field theory that arises from a composite fermion model.}
\end{figure}

Once the Grassmann integration is performed the partition function can be written as a statistical mechanics of two different types of closed oriented loops which form the world lines of the four pions in the model. A non-zero $c$ creates lattice sites which the pion loops cannot touch. An example of a configuration of pions is given in Fig.\ref{fPIconf}. We refer to such sites as ``instantons''. The model has been studied recently with a directed path algorithm and we refer the reader to the original work \cite{Cecile:2007dv}. Cluster algorithms in the conventional formulation do not work for scalar field theory with an $SU(2)\times SU(2)\times U(1)$ field theory. For this reason until recently it was difficult to study the effects of the anomaly on the two-flavor chiral transition. Using the worm-approach we were able to study this question non-perturbatively and beyond mean field theory for the first time and found that a rather strong anomaly was necessary before the $O(4)$ universality sets in \cite{Chandrasekharan:2007up}. Using this model we have also clearly shown that the $\sigma$-resonance can have a significant impact on the width of the region where chiral perturbation theory is valid \cite{Cecile:2008kp}. A variant of the above model also allows one to study an $SU(2)\times U(1)$ scalar field theory which naturally arises from two-color QCD with staggered fermions \cite{Chandrasekharan:2006tz}. 

Lattice QCD with staggered fermions at a finite baryon chemical potential in the strong coupling limit can also be formulated and studied in the WL-approach. This was already done almost two decades ago \cite{Karsch:1988zx}. The theory however suffers from a sign problem which can only be solved at zero chemical potential. In any case, in the previous study the problem was solved using a brute force technique but using a local algorithm. Recently, the problem has been revisited but now using the worm algorithm. Clear advantages of the worm approach has been demonstrated. For more details we refer to the work presented by Michael Fromm at this conference.

\section{World-line Approach: The Thirring Model}

Let us now discuss how the WL-approach can be applied to a fermionic field theory if the fermion sign problem can be solved. Solution to the fermion sign problem in three or more dimensions usually involves a re-summation over a class of configurations. This makes the problem computationally more demanding than bosonic field theories since a significant fraction of the update-time goes into the re-summation effort. Clearly, the new approach does not alleviate this problem. On the other hand it offers new techniques to solve the sign problem and some of them are better than the conventional methods. Conventionally, the solution to the sign problem in the presence of fermions comes from computing the determinant of a matrix whose size grows with system size. On the other hand in the WL-approach other interesting solutions which are computationally less demanding can emerge. One such approach called the meron-cluster approach was found about a decade ago \cite{PhysRevLett.83.3116}. In the following we will device a new determinantal solution to the lattice thirring model. The main advantage of this solution is that the size of the matrix is dynamically determined by the parameters of the model and does not always grow with system size. This is particularly true deep in the strong coupling phase (confined phase) which is the phase where the conventional algorithms become more demanding.

The action of the model we consider is constructed with massless staggered fermions and is given by
\begin{equation}
S = - \sum_{x,\mu} \eta_\mu(x) \psib_x[\psi_{x+\mu}-\psi_{x-\mu}] - U \psib_x\psi_x\psib_{x+\mu}\psi_{x+\mu}.
\label{tmod}
\end{equation}
Here we assume $x$ is a point on an $L^d$ hypercubic lattice with toroidal boundary conditions, the index $\mu=1,2,..,d$ labels the direction, the fermion fields $\psi_x$ and $\psib_x$ are single component Grassmann fields with anti-periodic boundary conditions. Since the interactions exist on bonds the model is equivalent to the massless Thirring model in the weak coupling limit \cite{DelDebbio:1995zc,Debbio:1997dv}. The lattice model is invariant under an exact $U(1)$ chiral symmetry:
\begin{equation}
\psi_x \rightarrow \exp(i\sigma_x\phi) \psi_x;\ \ \ \
\psib_x \rightarrow \exp(-i\sigma_x\phi) \psib_x
\end{equation}
where $\sigma_x$ is $1$ on even sites and $-1$ on odd sites. For small $U$ this chiral symmetry remains unbroken and the system contains massless fermions, while for large $U$ the symmetry is spontaneously broken and the system contains massless Goldstone bosons but massive fermions. Thus, there is an interesting quantum phase transition in the thermodynamic limit at a critical coupling $U_c$. A variant of this phase transition may be of interest in graphene-like systems \cite{hands-2008}.

The conventional Monte-Carlo approach is to rewrite the partition function
\begin{equation}
Z = \int \ [d\psib d\psi] \exp(-S)
\label{modpf}
\end{equation}
by introducing new auxiliary variables $\mathrm{e}^{i\phi_\mu(x)}$ on the links connecting $x$ with $x+\mu$ so that the four-fermion operator is converted to a fermion bilinear. Mathematically one can show that if we begin with
\begin{equation}
Z = \int \ d\phi \  [d\psib d\psi] \ \exp\Big\{\ 
\sum_{x,y} \psib_x (M[\phi])_{x,y} \psi_y \Big\}
\end{equation}
where $M(\phi)$ is an $L^d \times L^d$ matrix given by
\begin{equation}
M[\phi] = \eta_\mu(x) \Big[\delta_{x+\mu,y}(1+ \sqrt{U}\mathrm{e}^{i\phi_\mu(x)}) - \delta_{x-\mu,y}(1+\sqrt{U}\mathrm{e}^{-i\phi_\mu(x)})\Big],
\end{equation}
and integrate over the angles $0 \leq \phi_\mu(x) < 2\pi$ one can recover the original partition function given in eq.(\ref{modpf}). Bit if we integrate over the Grassmann variables first we obtain
\begin{equation}
Z = \int [d\phi] \ \mathrm{Det}(M(\phi))
\end{equation}
where the determinant is non-negative and thus the sign problem is solved. One can now use either the determinantal Monte-Carlo algorithm or the Hybrid Molecular Dynamics algorithm to generate $[\phi]$ accordingly \cite{DelDebbio:1995zc,Debbio:1997dv}. In the latter approach it is common to introduce a fermion mass in order to regulate the condition number of the fermion matrix $M[\phi]$. However, this breaks the chiral symmetry and it is difficult to extrapolate to the massless limit. The exactly massless problem of interest is known to be computationally very demanding especially for large values of $U$ due to the excessively large number of small eigenvalues of $M[\phi]$ \cite{hands-2008}. 

\begin{figure}[t]
\begin{center}
\includegraphics[width=0.7\textwidth]{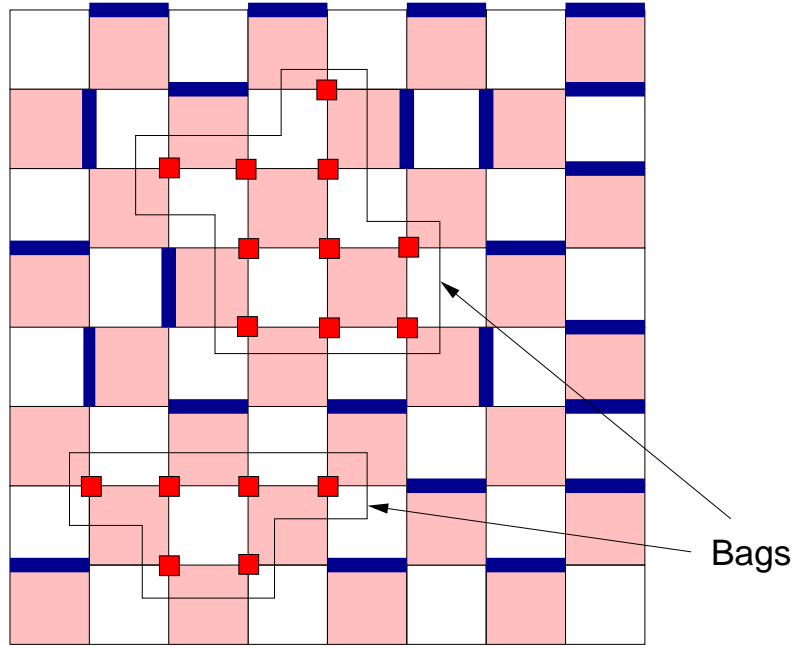}
\end{center}
\caption{\label{fig1} An illustration of a ``dynamical-bag'' configuration as discussed in the text.}
\end{figure}

The above approach is the most commonly used method to deal with four-fermion couplings and is known in the literature as a Hubbard-Stratanovich transformation. Let us now discuss a WL-approach to the problem. Instead of introducing auxiliary variables, we begin with the partition function given by
\begin{equation}
Z = \int [d\psi d\psib] \exp\Bigg(\sum_{x,\mu} 
\Big\{
\eta_\mu(x) \psib_x[\psi_{x+\mu}-\psi_{x-\mu}] + U \psib_x\psi_x\psib_{x+\mu}\psi_{x+\mu}\Big\}\Bigg),
\end{equation}
and expand it in powers of $U$ using
\begin{equation}
\exp(U \psib_x\psi_x\psib_{x+\mu}\psi_{x+\mu}) = 1 + U \psib_x\psi_x\psib_{x+\mu}\psi_{x+\mu}.
\end{equation}
The Grassmann integration then gives
\begin{equation}
Z = \sum_{n_{x,\mu}=0,1} \Big(\prod_{x,\mu} U^{n_{x,\mu}}\Big) \mathrm{Det}(W[n])
\end{equation}
where $n_{x,\mu}=0,1$. The $n_{x,\mu}=1$ bonds are the same as the hard core dimers encountered in the previous section in the fermionic $XY$ model. Note that in this approach the Grassmann integration leads to a determinant of a different matrix $W[n]$, which is just the free fermion matrix where the sites connected to $n_{x,\mu}=1$ are dropped. It is easy to argue that $\mathrm{Det}(W[n])$ is also non-negative and thus there is again no sign problem. 

Interestingly, the configuration $[n]$ divides the lattice into ``bags'' of sites connected with only $n_{x,\mu}=0$ bonds. Inside a bag the fermions hop freely while outside they are confined to hop on single bonds. The size and shape of the bags are dynamically determined by the value of $U$. Figure \ref{fig1} gives an illustration of one such configuration.  For this reason we call our new algorithm the ``dynamical-bag'' algorithm. Since a single world line configuration of fermions inside the bag can give negative weights due to the Pauli principle, one has to resum all the possible fermion world lines within the bag. This gives the $\mathrm{Det}(W[n])$, which in our case is non-negative. When $U$ is large, the bags are small comprising of a few neighboring bonds and thus independent of the system size. Hence, the computation of the determinant must be easy. In fact, as we will argue below, it is not even necessary to compute the determinant, but just one matrix element of the inverse of $W[n]$ at every local update step, which is quite easy for small matrices. Note that it is precisely when $U$ is large that conventional algorithms begin to fail. When $U$ is small the bags can percolate and become as big as the system size. This naturally leads to massless fermions. Here the WL-approach will be similar to the conventional approach in efficiency. The typical size of the bags may provide an interesting length scale for the problem. 

\begin{figure}[t]
\begin{center}
\includegraphics[width=0.7\textwidth]{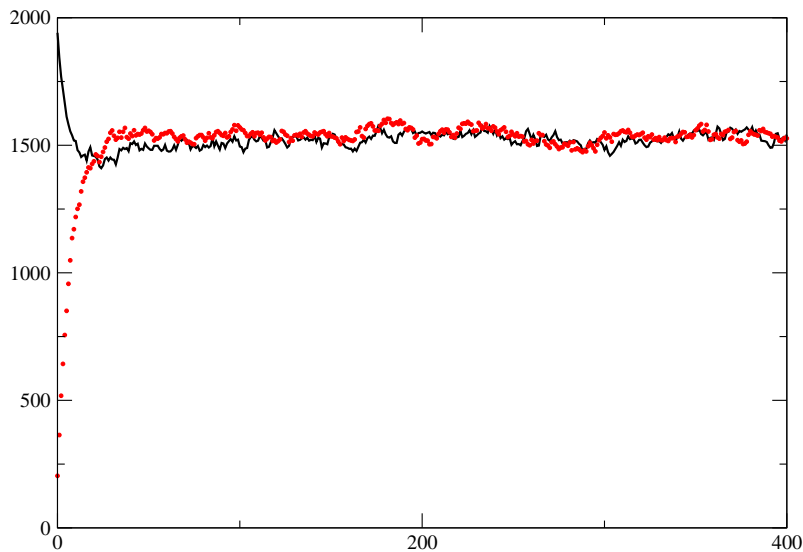}
\end{center}
\caption{\label{tevol} Plot of the number of $n_{x,\mu}=1$ bonds on a $16^3$ lattice at $U=1$ in the dynamical-bag algorithm.}
\end{figure}

Our algorithm consists of two steps: one step changes $n_{x,\mu}$ between $0$ and $1$ on the bonds and the other step moves the bonds $n_{x,\mu}$ bonds around. One can develop a combination of a regular local heat-bath algorithm to accomplish the former and a worm-type algorithm to accomplish the latter. An important point to note is that the probability to introduce or remove a bond depends on the ratio of two determinants one with and one without the bond. This ratio is exactly equal to $|[(W[n])^{-1}]_{x,x+\mu}|^2$ where the bond in consideration is between the sites $x$ and $x+\mu$ and the matrix $W[n]$ does not contain the bond under consideration. The computation of the inverse can be a time consuming process. Further, if the matrix $W[n]$ contains exact zero modes, one has to know about its existence so that such configurations are not generated. These two difficulties makes the algorithm more time consuming as compared to bosonic algorithms. However, this is a price one has to pay for doing fermionic physics and we do not know a way out of it at the moment. However, at large $U$ when the ``fermions'' are mostly bound into bosons these difficulties become milder and disappear completely at infinite $U$. Thus, the algorithm knows when to work hard to take into account the dynamics of the fermions. In Fig. \ref{tevol} we show the time evolution of the number of $n_{x,\mu}=1$ bonds on the lattice at $U=1$ starting from two different initial conditions on a $16^3$ lattice which was accomplished in a few hours on a laptop with just a local heat bath algorithm.

Although the local heat bath algorithm may be sufficient at small $U$, it will become inefficient as $U$ increases. At large $U$ the worm part of the algorithm is essential for efficiency. The worm algorithm is also necessary to measure observables such as the chiral condensate susceptibility which can get contributions from fermionic correlations between two different bags. These correlations are difficult to measure in the WL-approach using simple matrix inversions. In the worm algorithm such correlations come from configurations which contain two defects (the head and the tail of the worm) one of which is present in one bag and the other in the other bag. We are currently implementing this complete algorithm and the technical details will be discussed elsewhere.

\section{World-Sheet Approach: Abelian Gauge Theory}

The natural extension of world-lines to gauge theories must be in the form of world-sheets. Thus, the WL-approach must be modified into a WS-approach when applying to gauge theories. Here we illustrate the ideas using a pure compact $U(1)$ lattice gauge theory in four dimensions. In the conventional formulation the theory describes the dynamics of the compact gauge field $0 \leq \phi_\alpha(x) < 2\pi,\alpha=1,2,3,4$ which are angular variables associated to bonds $(x,\alpha)$ that connect sites $x$ and $x+\hat{\alpha}$. The model is described by the partition function
\begin{equation}
Z = \int [d\phi] \exp[- S].
\end{equation}
where the action is taken for convenience to be the Wilson action 
\begin{equation}
S = -\beta \sum_{P(x,\mu,\nu)} \cos\Bigg(\phi_{\mu\nu}(x)\Bigg),
\end{equation}
where the sum is over all plaquettes $P(x,\mu,\nu)$ (defined by a lattice site $x$ and two forward directions $\mu < \nu$) and $\phi_{\mu\nu}(x) = \phi_\mu(x) + \phi_\nu(x+a\hat{\mu}) - \phi_\mu(x+a\hat{\nu}) - \phi_\nu(x)$. We define $\phi_{-\alpha}(x) = - \phi_{\alpha}(x-\hat{\alpha})$ for convenience.

Motivated by the world-line approach we can rewrite the partition function in terms of world-sheet variables. For this purpose we begin with
\begin{equation}
Z = \int [d\phi] \prod_{P(x,\mu,\nu)} \mathrm{e}^{\beta \cos(\phi_{\mu\nu}(x))}.
\end{equation}
and use eq.(\ref{modbf}) on every plaquette and then integrate over all angles $\phi_\alpha(x)$ as in the $XY$ model. We now get
\begin{equation}
Z = \sum_{[k]} \prod_{P(x,\mu,\nu)} I_{k_{\mu\nu}(x)}(\beta) \ \prod_{(x,\mu)} \ 
\delta_{\sum_\nu [k_{\mu\nu}(x) - k_{\mu\nu}(x-\hat{\nu})],0}
\label{kpf}
\end{equation}
where $[k]$ represents a configuration of constrained integers on each plaquette $P(x,\mu,\nu)$. For simplicity we define $k_{\mu\nu}(x)=-k_{\nu\mu}(x)$. The configurations must now satisfy the constraint that
\begin{equation}
\sum_\nu [k_{\mu\nu}(x) - k_{\mu\nu}(x-\hat{\nu})] = 0
\end{equation}
at every bond $(x,\mu)$ and are shown with delta functions in the above expression. It is possible to argue that such a constrained set of integers arise from oriented closed surfaces or world sheets. For example, an update of $[k]$ requires the change of $k$ on a closed surface of plaquettes. Any configuration $[k]$ can be constructed by a series of local cubical updates and sheet updates in two dimensional planes.

An interesting observable is the Wilson loop. Consider a closed loop on a $\mu\nu$ plane of size $L \times L$. The Wilson loop associated with this loop is defined as 
\begin{equation}
W_L = \prod_{k=1}^N \exp(i\phi_{\alpha_k}(x_k))
\end{equation}
where $x_k$ and $\alpha_k$ are the sites and directions along the loop. The average of this observable is given by
\begin{equation}
\langle W_L \rangle  = \frac{1}{Z} \int [d\phi] \ W_L \ \exp\Big(-\beta S \Big) 
=  \Bigg\langle \prod_{p\in P} \frac{I_{k_p+1}(\beta)}{I_{k_p}(\beta)} \Bigg\rangle
\end{equation}
where the final average is assumed to be performed in the world-sheet representation and $P$ stands for the set of plaquettes that cover the surface of the Wilson loop.

\begin{figure}[t]
\vskip0.2in
\begin{center}
\includegraphics[width=0.7\textwidth]{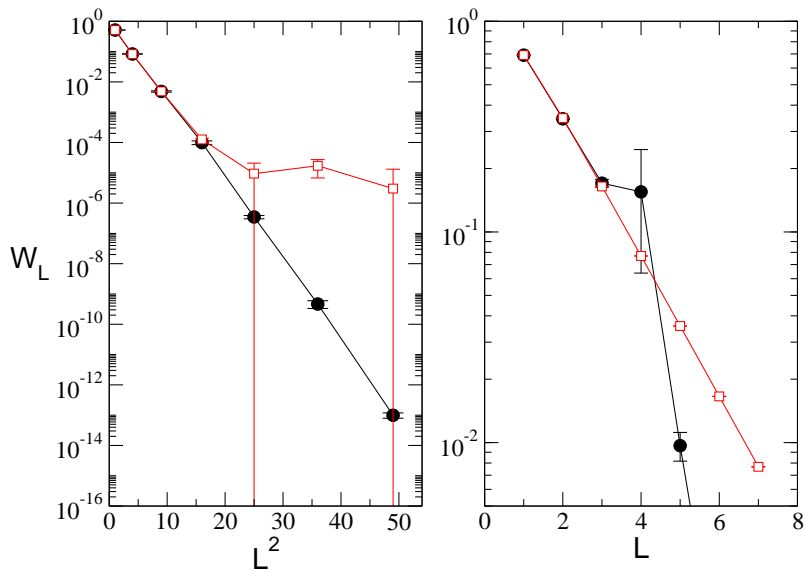}
\end{center}
\caption{\label{fig2} Comparing the results for the square Wilson-loop as a function of its size obtained using the world-sheet algorithm (filled circles) and the conventional algorithm (open squares). The world-sheet algorithm is better than the conventional method in the confined phase (left figure) while the opposite is true in the Coulomb phase (right figure).}
\end{figure}

In order to explore the usefulness of the new representation, we have implemented a worm-like update for the $[k]$ configuration which involves picking a $\mu\nu$ plane at random and introducing a defect in the world-sheet configuration in the form of a $1\times 1$ Wilson-loop. We then increase or decrease the size of the defect-loop but restricting it to be a square loop within the plane. During this process we allow for local cubical updates within a slice on either side of the sheet, which allows the surface being updated to fluctuate. The algorithm ends when the defect closes on itself or propagates and updates a whole sheet.We have observed that sheets do get updated in the Coulomb phase and when the volumes are small. Figure \ref{fig2} shows a comparison of the results of the world-sheet algorithm with the conventional algorithm for various sizes of a square shaped Wilson-loop. The figures show that for small sizes both algorithms are comparable and produce the expected physics, namely an area law in the confined phase and a perimeter law in the Coulomb phase. Interestingly, the world-sheet algorithm is able to measure large Wilson-loops quite well in the confined phase, while the conventional algorithm still appears superior in the Coulomb phase. 

\section{Conclusions}

We have tried to demonstrate in this talk that there is a new computational approach to a variety of lattice field theories. This approach is based on rewriting the partition function in a world-line representation. The constrained configurations in this representation can be updated efficiently with the newly discovered worm-algorithms. This approach gives an alternative method for studying many scalar field theories. They also lead to new solutions to the fermion sign problems. One such solution leads to an interesting fermion algorithm which we call the dynamical-bag algorithm. Finally we argue that the ideas may also be applicable to gauge theories in the form of world-sheet methods. Clearly the subject is still in its infancy and it would be exciting if some of the ideas can lead to important computational break-through.

\end{document}